\documentclass[%
 reprint,
 amsmath,amssymb,
 aps,
]{revtex4-1}

\usepackage{graphicx}
\usepackage{dcolumn}
\usepackage{bm}

\begin{document}

\preprint{APS/123-QED}

\title{Anomalous isotopic effect on
electron-directed reactivity
by a 3-$\mu$m midinfrared pulse}

\author{Kunlong Liu,$^{1}$ Qingbin Zhang,$^{1,2,*}$
Pengfei Lan,$^{1,2}$ and Peixiang Lu$^{1,2,}$}

\email{Corresponding authors: \\ zhangqingbin@mail.hust.edu.cn \\
 lupeixiang@mail.hust.edu.cn}

\affiliation{%
 $^1$Wuhan National Laboratory for Optoelectronics and School of Physics, Huazhong University of Science and Technology, Wuhan 430074, China \\
$^2$Key Laboratory of Fundamental Physical Quantities Measurement of Ministry of Education, Wuhan 430074, China
}%

\date{\today}

\begin{abstract}
We have theoretically studied the effect of nuclear mass on electron localization in dissociating H$_2$$^+$ and its isotopes subjected to a few-cycle 3-$\mu$m laser pulse.
Compared to the isotopic trend in the near-infrared regime, our results reveal an \textit{inverse} isotopic effect in which the degree of
electron-directed reactivity is even higher for heavier isotopes.
With the semi-classical analysis, we find, for the first time, the pronounced electron localization is established by the interferences through different channels of one- and, more importantly, higher-order photon coupling.
Interestingly, due to the enhanced high-order above-threshold dissociation of heavier isotopes, the interference maxima gradually become in phase with growing mass and ultimately lead to the anomalous isotopic behavior of the electron localization.
This indicates that the multi-photon coupling channels will play an important role in controlling the dissociation of larger molecules with midinfrared pulses.
\begin{description}
\item[PACS numbers]
33.20.Xx, 33.80.Wz, 32.80.Qk, 42.50.Hz
\end{description}
\end{abstract}

\pacs{Valid PACS appear here}
\maketitle

The electronic motion inside the molecule is of fundamental importance in determining the formation and fracture of chemical bonds. For more than two decades, many efforts have been done to study the electronic dynamics in laser-matter interactions \cite{Posthumus,Bandrauk2004,Roudnev,Zhou}, aiming at the control over ultrafast reactions.
With the recent development of laser techniques and attosecond science \cite{CorkumNP,Krausz,Lan},
it has become feasible to steer the electron localization in dissociating molecules with the carrier-envelope phase (CEP) stabilized few-cycle laser pulses \cite{Kling,Kremer,Tong,Grafe} or the sequential ultraviolet and near-infrared pulses \cite{He,Singh,Sansone}.
The asymmetric electron localization in molecules can be understood as the quantum interference of the populations that are resonantly transferred among at least two electronics states of different parity \cite{Ray,Fischer}.

More recently, the control of electron-directed reactivity in midinfrared laser pulses
has been explored to enhance the electron localization probability via the match between the duration of the few-cycle pulse and the dissociation time of the molecule \cite{IZm,Liu2,LanEL}.
To hold the control efficiency for further heavier nuclei, one may have to use the pulse with longer wavelength \cite{IZm}.
While this constitutes an important step towards the control of chemical reactions in larger molecules, we are wondering whether the physical mechanism responsible for the electron localization in midinfrared pulses remains the same as that in the near-infrared regime \cite{Kremer}.
Meanwhile, for extending the control scheme to the larger molecules, the influence of nuclear mass on the reactions must be taken into account. However, it remains unclear how the nuclear mass affects the control efficiency when the midinfrared pulses are applied, though it has been shown that the control of electron localization is weakened by the growing mass in the few-cycle near-infrared field \cite{Hua,Kremer}.

To understand the influence of mass in reactions, the use of isotopes is one of the practice means. This is because different isotopes of a given element have a similar electronic structure and chemical properties, but different masses. Indeed,
the isotopic effect has been an important issue in the research of laser-molecule interactions, such as high-harmonic generation \cite{Lein}, nuclear dynamics detection \cite{Baker} and high-order above-threshold dissociation (ATD) \cite{McKenna}.

On the other hand, thanks to the recent advance in the generation of midinfrared source, few-cycle CEP-stable pulses at 3 $\mu$m have become available in laboratory \cite{Biegert,Biegert2}, providing a potential for controlling as well as obtaining deeper insight into the electron localization process in heavier molecules.
In this Letter, we theoretically study the asymmetric dissociation of molecular hydrogen by using a few-cycle 3-$\mu$m pulse
and
report an anomalous isotopic effect on the electron-directed reactivity.
The dissociation process is analyzed with the semi-classical model \cite{Freek}. For the 3-$\mu$m field, the underlying mechanism for the electron localization is found to be
based on the interferences between the one- and higher-order photon coupling channels.
Due to the
enhanced ATD for heavier isotopes \cite{McKenna},
the interference maxima gradually become in phase with growing nuclear mass, resulting in the anomalous isotopic effect where the electron localization degree is larger for heavier isotopes.

In our simulations, we have used a numerical model that has been well established for studies of electron localization in the laser-driven molecular dissociation \cite{Kling,Kremer}. This model solves the time-dependent Schr\"{o}dinger equation (TDSE) for the nuclear wavepacket in the Born-Oppenheimer (BO) representation.
In the present model, we have considered the molecular ions oriented along the polarization axis. Since the time scale of the nuclear rotation is several hundred femtoseconds,
two reasonable assumptions underlie our calculation \cite{Wang}: (i) the nuclei do not have time to rotate during the time of interaction and (ii) the nuclear rotation after the pulse is not significant.
The molecular dynamics of H$_2$$^+$ can be well described in terms of the two lowest-lying electronic states, the $1s\sigma_g$ and $2p\sigma_u$ states. The TDSE can be written as (atomic units are used throughout unless otherwise indicated) \cite{Kling}:
\begin{eqnarray}
i\frac{\partial}{\partial t}
\begin{pmatrix}
\chi_g \\
\chi_u
\end{pmatrix}
&=&
\begin{pmatrix}
-\frac{\nabla ^2}{2\mu}+V_g & V_{gu} \\
V_{gu}^* & -\frac{\nabla ^2}{2\mu}+V_u
\end{pmatrix}
\times
\begin{pmatrix}
\chi_g \\
\chi_u
\end{pmatrix},
\end{eqnarray}
where $\mu$ is the nuclear reduced mass, and $V_g$ and $V_u$ are the BO potential curves of the $1s\sigma_g$ and $2p\sigma_u$ states, respectively.
The coupling term $V_{gu}$ equals $E(t)\times D_{gu}$, with $D_{gu}$ the electronic transition dipole between the two states. The electric field $E(t)$ of the pulse is expressed as
$
E(t)=\varepsilon_0\exp[-2\ln2(t/\tau)^2]\cos(\omega t+\phi),
$
where $\varepsilon_0$, $\tau$, $\omega$, and $\phi$ denote the electric field intensity, pulse duration ($\mathrm{FWHM}$), frequency, and CEP of the pulse, respectively.

The time-dependent nuclear wave functions corresponding to the `left' ($l$) and `right' ($r$) atoms within the molecule are defined as
$\Psi_{l,r}=(\chi_g\pm\chi_u)/\sqrt{2}$.
The kinetic energy release (KER) spectrum $S_{l,r}(E_k)$ is obtained by Fourier analysis of the real-space wave function with the bound states projected out \cite{Abeln}.
The final probabilities $P_{l,r}$ of the electron being localized on the left or right nucleus are calculated by integrating $S_{l,r}(E_k)$ over the kinetic energy $E_k$. In respect that the preparation mechanism of H$_2$$^+$ generally indicates an incoherent Franck-Condon (FC) distribution of vibrational states in experiments \cite{McK},
we have averaged the observables over the initial vibrational states weighted by FC factors in the present calculation.

\begin{figure}[b]
\centering\includegraphics[width=8.2cm]{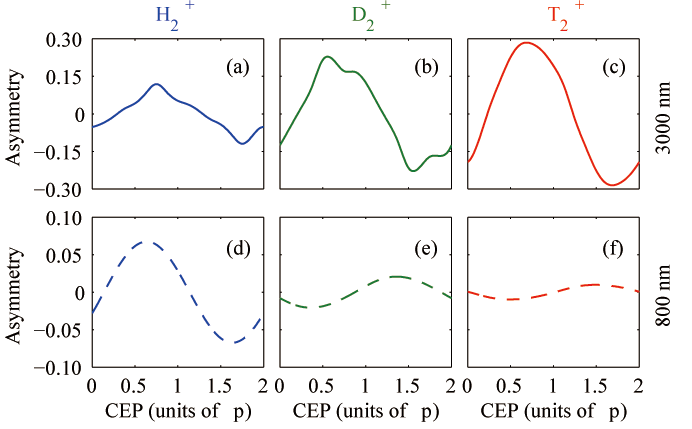}
\caption{\label{fig1}
The CEP dependence of the FC-averaged electron localization asymmetry of H$_2$$^+$ and its isotopes, D$_2$$^+$ and T$_2$$^+$, for the pulse wavelengths of 3000 (upper row) and 800 nm (lower row).
}
\end{figure}
%

We study the isotopic effect on the asymmetric dissociation of H$_2$$^+$ with a three-cycle (FWHM), 3000-nm laser pulse with an intensity of $
1\times 10^{14}\ \mathrm{W/cm}^2$. The asymmetry parameter of the final electron localization probability is defined as $A=(P_l-P_r)/(P_l+P_r)$.
Figures~\ref{fig1}(a)--\ref{fig1}(c) reveal the asymmetries after FC averaging as a function of CEP in the 3000-nm pulse
for H$_2$$^+$ and its isotopes, respectively.
For comparison, a similar simulation is performed for a three-cycle, 800-nm pulse with the same pulse intensity. The results are shown in Figs.~\ref{fig1}(d)--\ref{fig1}(f).
First of all, it is obvious that the asymmetry is enlarged in the midinfrared field for each isotope, as predicted in \cite{IZm,Liu2}.
Then, more interestingly, we can obtain two contrastive observations from Fig.~\ref{fig1}. (i) The asymmetry for the 800-nm pulse displays a CEP dependence that decreases with growing mass. This is coincident with the results of the previous studies \cite{Hua}.
But curiously, the asymmetry for the 3000-nm pulse shows an inverse isotopic behavior in which the degree of the asymmetry is even higher for heavier isotopes. (ii) The curves of the asymmetry for the 800-nm pulse show a regular sine form, but they become irregular for the 3000-nm pulse.

\begin{figure}[b]
\centering\includegraphics[width=8.5cm]{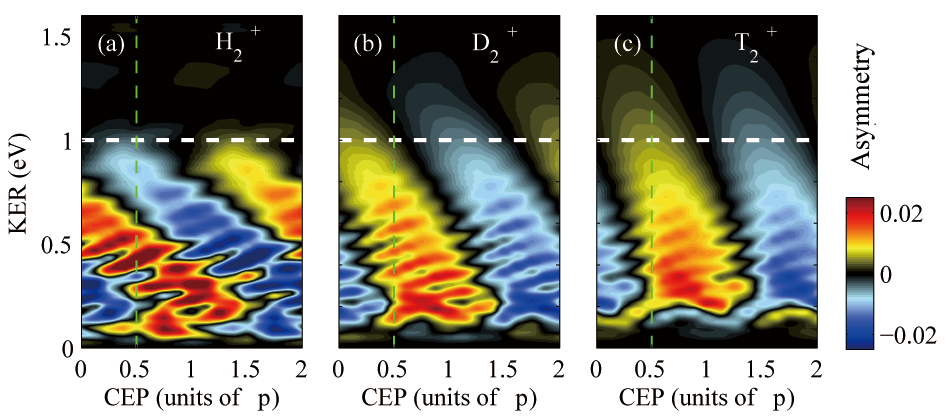}
\caption{\label{fig2}
The asymmetry parameter $A(E_k,\phi)$ as a function of the CEP and the kinetic energy for (a) H$_2$$^+$, (b) D$_2$$^+$, and (c) T$_2$$^+$ in the 3000-nm pulse. The other laser parameters are the same as in Figs.~\ref{fig1}(a)--\ref{fig1}(c)
}
\end{figure}

To gain insight into the interaction dynamics under the 3-$\mu$m field, we
present the absolute asymmetry as a function of the CEP and the kinetic energy in Fig.~\ref{fig2},
with the same laser parameters as in Figs.~\ref{fig1}(a)--\ref{fig1}(c).
The absolute asymmetry is given by $A(E_k,\phi)=S_{l}(E_k,\phi)-S_{r}(E_k,\phi)$.
Generally, the asymmetry shows a strong dependence on both the CEP and the KER for all isotopes.
In detail, we see, on one hand, two major ``undulate tilted stripes'' corresponding to the left-right asymmetry.
As the nuclear mass increases, the stripes gradually become more and more precipitous and they are almost vertical for T$_2$$^+$.
Additionally, as indicated by the horizontal dashed line, the asymmetry is almost stop at about 1 eV for H$_2$$^+$, but the strong asymmetry extends to higher energy with growing mass.
Consequently, after the integral over the KER, the degree of the asymmetry turns out to be larger for the heavier isotope.
On the other hand, we can see that the stripes of the spectra contain several peaks.
In the case of H$_2$$^+$ in Fig.~\ref{fig2}(a), the peaks at different KER respond inconsistently to the CEP, leading to the irregular asymmetry curve after the KER integral, as shown in Fig.~\ref{fig1}(a). But, for T$_2$$^+$ in Fig.~\ref{fig2}(c), the alignment of the peaks are almost in step. Therefore, the asymmetry for T$_2$$^+$ in Fig.~\ref{fig1}(c) becomes closer to a sine-like curve.

While the KER distributions give us a well understanding of the features of the CEP-dependent asymmetry in Fig.~\ref{fig1}, the physical mechanism for the anomalous isotopic behavior of the electron localization has remained unclear: Why are there asymmetry peaks at different kinetic energy and why do those peaks gradually become in step with growing nuclear mass?
In order to uncover the underlying dynamical process, we analyze the molecular dynamics with the semi-classical model
in terms of the quasi-static states \cite{Freek}, which are related to the left and right localized wave functions $\Psi_{l,r}$ via
$\psi_{1,2}=[(\cos\theta\pm\sin\theta)\Psi_{l}
\pm(\cos\theta\mp\sin\theta)\Psi_{r}]/\sqrt{2}$,
with $2\theta=\tan^{-1}[-2V_{gu}/(V_u-V_g)]$.
The corresponding quasi-static eigenvalues can be obtained by
\begin{eqnarray}
V_{1,2}=\frac{V_g+V_u}{2} \mp
\sqrt{\frac{(V_u-V_g)^2}{4}+V_{gu}^{2}}\ .
\end{eqnarray}
It has been demonstrated that the electron localization is established by the passage of the dissociating molecule through an intermediate regime where the lase-molecule interaction is neither diabatic nor adiabatic \cite{Freek}.
This process can be evaluated by the transition probability
\begin{eqnarray}
P_{ts}(t)=\exp\{-\frac{\pi[V_u(R)-
V_g(R)]^2}{4\omega D_{gu}(R)E_{0}(t)}\},
\end{eqnarray}
where $\omega$ is the laser frequency mentioned above, $E_{0}(t)$ is the envelope of the laser field, and $R$ tracks the expectation value of the internuclear distance of the time-dependent nuclear wavepacket.
Then we choose the molecular ions H$_2$$^+$ of the 5th vibrational state ($v=5$) and D$_2$$^+$ of the 7th vibrational state ($v=7$) as the investigation targets. The initial equilibrium distances (2.634 and 2.641 a.u.) and the vibrational state energies ($-0.5608$ and $-0.5603$ a.u.) of them are close to each other so that we can focus on the influence of the nuclear mass on the dissociation process.

\begin{figure}[b]
\centering\includegraphics[width=8.2cm]{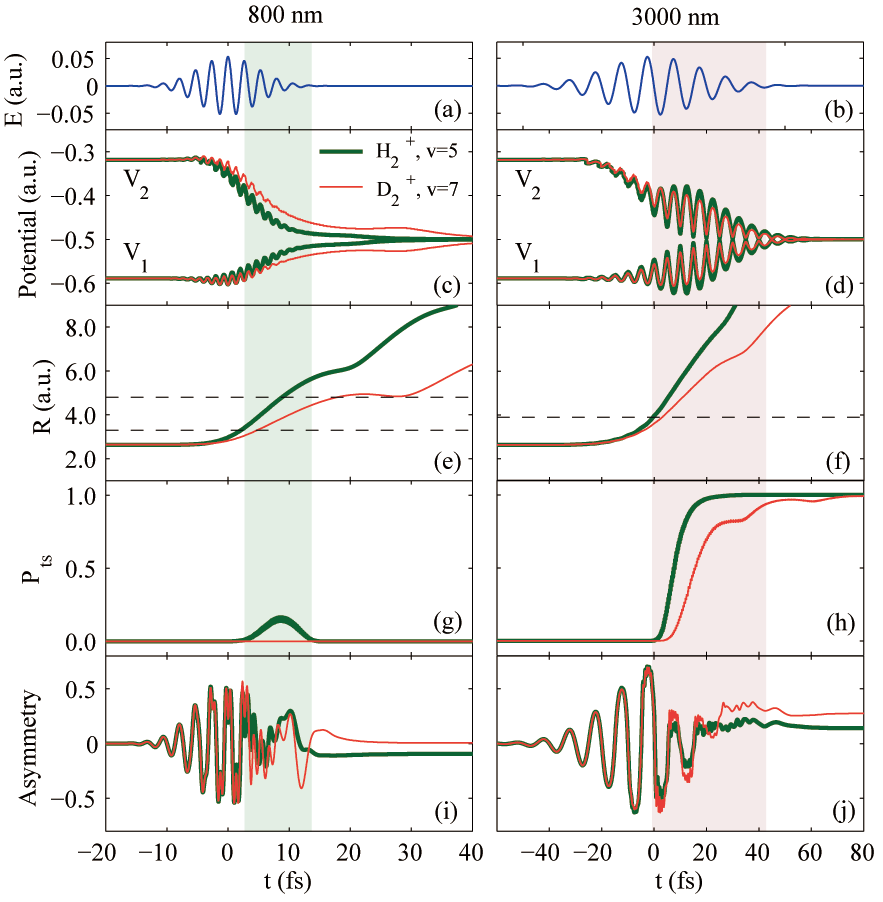}
\caption{\label{fig3}
The time evolution of the physical quantities of the electron localization process in the 800- and 3000-nm fields for H$_2$$^+$ (thick curve) and D$_2$$^+$ (thin curve). (a)--(b) The electric field. (c)--(d) The quasi-state eigenvalues. (e)--(f) The internuclear distance. (g)--(h) The transition probability. (i)--(j) The localization asymmetry. The shadows indicate the regions where the electron localization is established.
}
\end{figure}

In Fig.~\ref{fig3}, from up to down the left and right columns depict the time evolution of the electric field $E$,
the quasi-state eigenvalues $V_{1,2}$,
the internuclear distance $R$, the transition probability $P_{ts}$, and the localization asymmetry for the 800- and 3000-nm pulses, respectively.
According to the semi-classical model,
the electron localization is established though the region where $0<P_{ts}<1$, as shown by the shadows in the figure.
We find that the evolutions of $P_{ts}$ in the 3000-nm pulse [Fig.~\ref{fig3}(h)] are quite different from those in the 800-nm pulse [Fig.~\ref{fig3}(g)]. This indicates a distinct physical mechanism responsible for the electron localization in the 3000-nm field.

In the case of the near-infrared pulse (the left column of Fig.~\ref{fig3}), the nuclei separate slowly [Fig.~\ref{fig3}(e)].
At the early time of the interaction, the energy gap between $V_{1,2}$ is large and the dynamics is adiabatic. As the interaction continues, the energy gap becomes smaller.
For H$_2$$^+$ (thick solid curve), the dynamics becomes non-adiabatic when the internuclear distance sequentially passes through the three- and one-photon crossings, as indicated by the dashed lines in Fig.~\ref{fig3}(c).
This can be understood as that the electron localization is established
by interference between the dissociative populations that are generated through the three- and one-photon coupling channels.
After that,
the electric field is no longer strong enough to overcome the energy gap and the dynamics becomes adiabatic again.
For D$_2$$^+$ (thin solid curve), however, due to the slower nuclear motion, the pulse becomes much weaker when the internuclear distance passes the coupling region.
Meanwhile, the energy gap is still large for D$_2$$^+$ [Fig~\ref{fig3}(c)], so the transition probability is near upon zero, resulting in the small asymmetry.
Therefore, under the mechanism of the near-infrared field, the electron localization is more difficult to be achieved for heavier molecules.

However, the underlying dynamics is different in the midinfrared field (the left column of Fig.~\ref{fig3}).
As shown in Fig.~\ref{fig3}(d),
the evolutions of $V_{1,2}$ are almost the same for both molecular ions.
According to the transition probability in Fig.~\ref{fig3}(h), the dynamics becomes non-adiabatic when the internuclear distances pass about $R=3.9$ a.u. [corresponding to the nine-photon crossing of 3000 nm, indicated by the dashed line in Fig.~\ref{fig3}(f)].
Then the electron localizations for both molecules begin to be established at almost the same time.
Due to the fast separation of the nuclei, the dynamics has been diabatic when the pulse is turned off and, consequently, the electron localization is frozen.
Similarly to the situation in 800-nm field, the effect of nuclear mass in the 3000-nm field is to slow the nuclear motion.
But the midinfrared pulse is strong and long enough to overcome the energy gap. Moreover, a distinct difference is that
the higher-order multi-photon crossing is open in the midinfrared field. The population is therefore transferred between the gerade and ungerade states through those multi-photon coupling channels.
%
Thus the pronounced electron localization can still be achieved for heavier isotopes.

%

\begin{figure}
\centering\includegraphics[width=8.3cm]{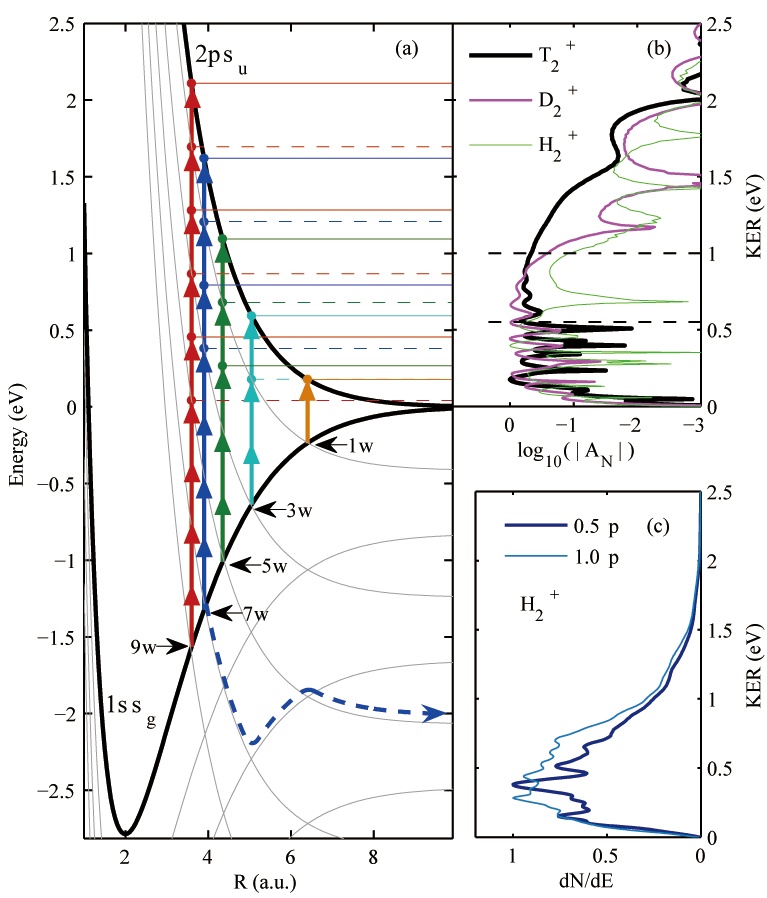}
\caption{\label{fig4}
(a) Illustration of the dissociation model for H$_2$$^+$ in the 3000-nm field. The vertical lines with arrows reveal the coupling channels that are noted by the text-arrows. The thin gray curves indicate the BO potentials dressed by net absorbed numbers of photons. The blue dashed arrow shows one of the possible dissociation pathways. (b) The absolute value of a cut of the asymmetry $A(E_k,\phi)$ normalized to the maximum asymmetry at the given CEP of $\phi=0.5\pi$, as marked by the vertical dashed lines in Fig.~\ref{fig2}. (c) The KER spectra normalized to the maximum value for the H$_2$$^+$ dissociation under the 3000-nm pulse with the CEP of $0.5\pi$ and $1.0\pi$.
}
\end{figure}

Based on the semi-classical analysis above, next we will introduce the dissociation model illustrated in Fig.~\ref{fig4}(a) and then discuss the reason for the anomalous isotopic effect in the 3000-nm field.
The dissociation process can be understood as follows.
During the interaction, the outgoing wavepacket sequentially passes through the $9\omega$ to $1\omega$ crossings shown by the text-arrows in Fig.~\ref{fig4}(a). Then, the populations are resonantly transferred between the $1s\sigma_g$ and $2p\sigma_u$ states via absorption or emission of photons at the crossing.
Note that for the few-cycle pulses, some of the crossings may be ``switched off'' and the dissociation pathways would be changed if the CEP is varied. This is because the instantaneous electric field may be zero when the wavepacket reaches the crossing.
Therefore, the molecule could finally dissociate via net-absorption of one or more photons at different initial coupling channels.
For example, the dashed arrow in Fig.~\ref{fig4}(a) depicts one of the possible pathways that start at the $7\omega$ coupling channel: The wavepacket first absorbs seven photons at the $7\omega$ crossing, then passes ``$5\omega$'' without emission of photons but emits three photon at ``$3\omega$'', and finally reabsorbs one photon at ``$1\omega$'', leading to the net five-photon absorption at the initial crossing of ``$7\omega$''.

In Fig.~\ref{fig4}(a), the centers of the KER distribution for different dissociation pathways have been shown by the horizontal dashed or solid lines.
We also show the FC-averaged KER spectra normalized to the maximum value for the H$_2$$^+$ dissociation under the 3000-nm pulse with the CEP of $0.5\pi$ and $1.0\pi$
in Fig.~\ref{fig4}(c). Both spectra range from 0 to 2.5 eV and involve several peaks that originate from the different dissociation pathways. The change of the peak values with the CEP demonstrates that the contributions from different pathways are indeed affected by the CEP as mentioned above.

Note that those horizontal dashed and solid lines in Fig.~\ref{fig4}(a) suggest the dissociative populations on the $1s\sigma_g$ and the $2p\sigma_u$ states, respectively. Due to the large laser bandwidth of the pulse, the KER distributions for the pathways of different parity overlap in energy, leading to the interferences and thus the peaks of the asymmetry at different KER.
By changing the CEP of the pulse, the contributions from different pathways will be modulated, so that the electron localization exhibits the
CEP- as well as KER-dependent asymmetry shown in Fig.~\ref{fig2}.

For comparison, in Fig.~\ref{fig4}(b) we present the asymmetry parameter $A_N$ as a function of kinetic energy for three isotopes,
where $A_N$ is the absolute value of a cut of the asymmetry $A(E_k,\phi)$ normalized to the maximum asymmetry at $\phi=0.5\pi$, as marked by the vertical dashed lines in Fig.~\ref{fig2}.
One can see that all the coupling channels, from ``$9\omega$'' to ``$1\omega$'', contribute to the asymmetry at lower KER between 0--0.55 eV. But the asymmetry of higher KER is mainly contributed from the relative higher-order coupling channels.
According to the semi-classical analysis, the strength of the relative lower-order coupling channels will be shrunk for heavier isotopes due to slower motion of the wavepacket which reaches the coupling region when the electric field is weaker.
Relatively,
the contributions of the higher-order channels will become more important to the dissociation of heavier isotopes.
This process can also be understood as the enhanced high-order ATD \cite{McKenna}.
Therefore, the asymmetry is gradually enhanced in higher energy regions with growing mass, as indicated by the horizontal dashed lines in Fig.~\ref{fig4}(b) and Fig.~\ref{fig2}.

Furthermore, the phase shift of the asymmetry peaks at different KER depends on the relative phase between the overlapped populations on the gerade and ungerade states \cite{Fischer}.
For the dissociation pathways starting at different initial crossings, the corresponding dissociative wavepackets have the different initial phases and thus respond inconsistently to the CEP.
Consequently, the phase of the CEP-dependent asymmetry shifts with the KER, resulting in the tilted asymmetry stripes shown in Fig.~\ref{fig2}.
But due to the relative enhancement of the higher-order coupling channels for heavier isotopes, the CEP response of the asymmetry at higher KER gradually becomes consistent to that at lower KER.
Therefore,
the alignment of the asymmetry peaks is more vertical (KER-independent) for heavier isotopes, ultimately leading to the isotopic behavior where the degree of electron localization increases with growing nuclear mass.

In conclusion, we have studied the electron localization in H$_2$$^+$ and its isotopes by the few-cycle 3-$\mu$m pulse and have found an anomalous isotopic effect on the electron-directed reactivity.
Compared to the situation in near-infrared fields, a distinct underlying dynamics in the midinfrared pulses is the opening of the higher-order (more than three) multi-photon coupling channels.
Pronounced electron localization, even for heavier isotopes, can be achieved through interferences between the dissociation pathways that start at the high-order crossings.
As the nuclear mass increases,
the contribution of the high-order coupling channels is enhanced and the interference maxima gradually become in step, resulting in the larger dissociation asymmetry for heavier isotopes.
This unexpected isotopic behavior has provided new insights into the electronic dynamics inside molecules. In the future,
the high-order multi-photon coupling channels will play
an important role in the control over electron-directed reactivity of larger molecules with midinfrared pulses.
%

%

This work was supported by the NNSF of China
under Grants
No. 60925021, No. 10904045 and No. 11234004, the 973 Program of China under Grant No.
2011CB808103, and the Doctoral Fund of Ministry of Education of China under Grant No.
20100142110047.

\end{document}